\documentclass[journal]{IEEEtran}

\usepackage{subfigure}
\usepackage{amsmath,amssymb}
\usepackage[dvips]{graphicx}
\usepackage{amsfonts}
\usepackage[mathscr]{eucal}
\usepackage{latexsym}
\usepackage{amsthm}
\usepackage{exscale}
\usepackage[mathscr]{eucal}
\usepackage{bm}
\usepackage[dvipsnames]{color}
\usepackage{cases}
\usepackage{epsfig}
\usepackage[center,small]{caption}
\usepackage{algorithm}
\usepackage{algorithmic}
\usepackage[verbose,nospace,sort]{cite}
\usepackage{tabularx}
\usepackage{multirow}
\usepackage{multicol}
\usepackage{balance}

\graphicspath{{./figs/}}

\scrollmode

\newtheorem{theorem}{Theorem}

\newtheorem{lemma}{Lemma}

\newcommand{\Sn}{\mathrm{S}}
\newcommand{\D}{\mathrm{D}}
\newcommand{\R}{\mathrm{R}}
\newcommand{\SR}{\mathrm{sr}}
\newcommand{\RD}{\mathrm{rd}}

\newcommand{\opt}{*}

\begin{document}
\title{Throughput Maximization for Mobile Relaying Systems}
\author{Yong~Zeng, Rui~Zhang, and Teng Joon Lim \\
Department of Electrical and Computer Engineering, National University of Singapore\\
 e-mail: \{elezeng, elezhang, eleltj\}@nus.edu.sg
\vspace{-3ex}
}

\maketitle

\begin{abstract}
This paper studies a novel \emph{ mobile relaying} technique, where relays of high mobility are employed to assist the communications from source to destination. 
By exploiting the predictable channel variations  introduced by relay mobility, we study the throughput maximization problem in a mobile relaying system via dynamic rate and power allocations at the source and relay. An optimization problem is formulated for a finite time horizon, subject to an \emph{information-causality constraint}, which results from the data buffering employed at the relay. It is found that the optimal power allocations across the different time slots follow a ``stair-case'' water filling (WF) structure, with \emph{non-increasing} and \emph{non-decreasing} water levels at the source and relay, respectively. For the special case where the relay moves unidirectionally from source to destination, the optimal power allocations reduce to the conventional WF with constant water levels. Numerical results show that with appropriate trajectory design, mobile relaying is able to achieve tremendous throughput gain over the conventional static relaying.
\end{abstract}

\section{Introduction}
In wireless communication systems, relaying is an effective technique for throughput/reliability improvement as well as range extension \cite{23},\cite{634}. 
 However, due to the practical  constraints such as limited node mobility and wired backhauls, most of the existing relaying techniques are based on relays deployed  in fixed locations, or {\it static relaying}. 
In this paper, we propose a novel relaying technique, termed {\it mobile relaying}, where the relay nodes are assumed to be capable of moving at relatively high speed, e.g., enabled by terminals mounted on ground or aerial vehicles. We note that the practical deployment of dedicated mobile relaying nodes is becoming more feasible than ever before, thanks to the continuous cost reduction in autonomous or semi-autonomous vehicles, such as unmanned aerial vehicles (UAVs) \cite{616}, as well as drastic device miniaturization in communication equipment. Compared with the conventional static relaying, 
the high mobility of mobile relays offers new opportunities for performance enhancement through the dynamic adjustment of relay locations to best suit the communication requirement, a technique that is especially promising for delay-tolerant applications such as periodic sensing. Note that while node mobility has been well exploited for upper layer designs in communication networks \cite{638}, to the best of our knowledge, its exploitation for physical layer designs is still under-developed.

By exploiting the predictable channel variations introduced by relay mobility along fixed paths, we study the throughput maximization problem via dynamic rate and power allocations at the source and relay. Unlike the conventional static relaying schemes \cite{639},\cite{640}, we employ a so-called {\it decode-store-and-forward} (DSF) strategy for the proposed mobile relaying, where, if necessary, the data received by the relay is temporarily stored in a data buffer before being forwarded to the destination. A throughput maximization problem is formulated for a finite time horizon subject to a new {\it information-causality constraint}, i.e., the relay can only forward the data that has already been received from the source over the previous time slots. We show that the optimal power allocations across different slots follow a ``stair-case'' water-filling (WF) structure  in general, with {\it non-increasing} and {\it non-decreasing} water levels at the source and relay, respectively. It is interesting to note that such a result is analogous to the power allocation in {\it energy harvesting communications} \cite{636,635,637}.
It appears that causality constraints, whether information or energy causality, induces a directional water filling optimal power allocation.
For the special case where the relay node moves unidirectionally towards the destination, we obtain the optimal solution in closed-form.

It is worth remarking that unlike the existing {\it buffer-aided} static relaying technique \cite{625}, which mainly relies on random channel fading for opportunistic link selections for throughput enhancement, the proposed mobile relaying in fact pro-actively constructs favorable channel conditions via careful mobility control, and thus introduces an additional degree of freedom for performance enhancement.

\begin{figure}
\centering
\includegraphics[scale=0.5]{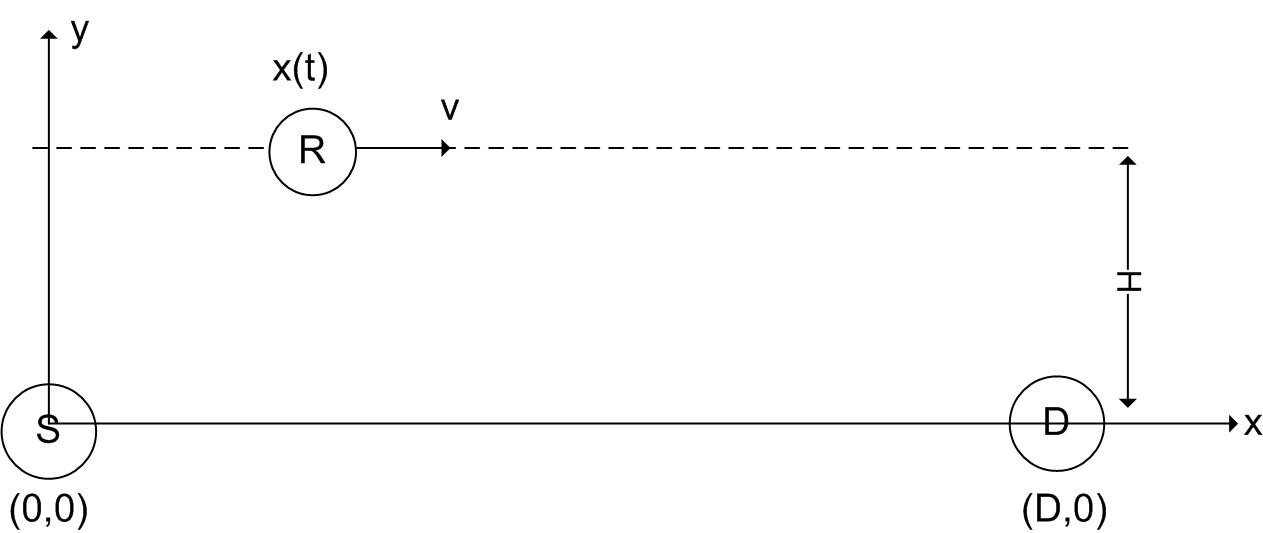}
\caption{A mobile relaying system.\vspace{-3ex}}\label{F:MobileRelay}
\end{figure}

\section{System Model and Problem Formulation}
As shown in Fig.~\ref{F:MobileRelay}, we consider a wireless system with a source node $\Sn$ and a destination node $\D$ which are separated by $D$ meters. We assume that the direct link between $\Sn$ and $\D$ is negligible due to e.g., severe blockage. Thus, a relay $\R$ needs to be deployed to assist the communication from $\Sn$ to $\D$. Unlike the conventional static relaying techniques with fixed relay locations,  we assume that a relay of high mobility is employed. In the following, we focus on UAV-enabled mobile relaying, but the design principles are applicable for the generic mobile relaying techniques.

 We consider a two-dimensional (2D) coordinate system with $\Sn$ and $\D$ located at $(0,0)$ and $(D,0)$, respectively, as shown in Fig.~\ref{F:MobileRelay}. We assume that a UAV  flying at a constant altitude $H$ is employed as a mobile relay for a finite time horizon $T$. Thus, the time-varying coordinate of the relay node $\R$ can be expressed as $(x(t),H)$, $0 \leq t \leq T$, with $x(t)$ denoting the relay's x-coordinate. We assume that $0\leq x(t) \leq D$, $\forall t$, i.e., the relay is always located in between the source and the destination.  Denote the maximum UAV speed as $\tilde{V}$. We thus have $|\dot{x}(t)|\leq \tilde{V}$, $0\leq t\leq T$, with $\dot{x}(t)$ denoting the time-derivative of $x(t)$. For ease of exposition, the time horizon $T$ is discretized into $N$ equally spaced time slots, i.e., $T=N\delta t$, with $\delta t$ denoting the elemental slot length, which is chosen to be sufficiently small so that the UAV's location can be assumed to be constant within each slot. Thus, the UAV's trajectory $x(t)$ can be approximated by the $N$-length sequence $\{x[n]\}_{n=1}^N$, where $x[n]$ denotes the UAV's x-coordinate at slot $n$. Furthermore, the speed constraint  can be written as
$|x[n+1]-x[n]|\leq \tilde{V}\delta t \triangleq V, \ n=1, \cdots, N-1$.

  For simplicity, we assume that $\R$ is equipped with a data buffer of sufficiently large size, and it operates in a full-duplex mode with concurrent information reception from $\Sn$ and transmission to $\D$ with perfect self-interference cancelation \cite{612}. 
 For ease of exposition, we assume that the communication from $\Sn$ to $\R$ and that from $\R$ to $\D$ are dominated by line-of-sight (LoS) links. Furthermore, the Doppler effect due to the relay's mobility is assumed to be perfectly compensated. Thus, at slot $n$, the channel power from $\Sn$ to $\R$ follows the free-space path loss model as
 \begin{align}
 h_{\SR}[n]=\beta_{0} d_{\SR}^{-2}[n]=\frac{\beta_{0}}{H^2 + x^2[n]},  \ n=1,\cdots, N,
 \end{align}
 where $\beta_{0}$ denotes the channel power at the reference distance $d_0=1$ meter, whose value depends on the carrier frequency, antenna gain, etc., and  $d_{\SR}[n]=\sqrt{H^2+ x^2[n]}$ is the link distance between $\Sn$ and $\R$ at slot $n$. Let $p_s[n]$ denote the transmission power by $\Sn$ at slot $n$. The maximum transmission rate by $\Sn$ to $\R$ in bits/second/Hz (bps/Hz) for slot $n$ can then be expressed as
 \begin{align}
R_{s}[n]&= \log_2\left(1+\frac{p_s[n] h_{\SR}[n]}{\sigma^2}\right),\\ 
&= \log_2\left(1+\frac{p_s[n] \gamma_{0}}{H^2+ x^2[n]}\right), \ n=1,\cdots, N,\label{eq:RSR}
 \end{align}
 where $\sigma^2$ denotes the noise power, and $\gamma_{0}\triangleq \beta_{0}/\sigma^2$ represents the reference signal-to-noise ratio (SNR). Similarly,  the channel from $\R$ to $\D$ at slot $n$ can be expressed as
$h_{\RD}[n]= \beta_{0}/(H^2+(D-x[n])^2)$, and the maximum transmission rate by $\R$ is
 \begin{align}
 R_{r}[n]&=\log_2\big(1+\frac{p_r[n]\gamma_{0}}{H^2+ (D-x[n])^2}\big), n=1,\cdots, N,
\end{align}
where $p_r[n]$ represents the transmission power by $\R$ at slot $n$.

Moreover, at each slot $n$, $\R$ can only forward the data that has already been received from $\Sn$. By assuming that the processing delay at $\R$ is one slot, we have the following {\it information-causality constraint} \cite{635}
\begin{align}
R_r[1]=0, \ \sum_{i=2}^n R_r[i] \leq \sum_{i=1}^{n-1} R_s[i], n=2,\cdots, N.
\end{align}
It is not difficult to see that $\Sn$ should not transmit at the last slot $N$. We thus have $R_s[N]=R_r[1]=0$, and hence $p_s[N]=p_r[1]=0$.
For a given UAV trajectory $\{x[n]\}_{n=1}^N$, define the time-dependent channels for the $\Sn$-$\R$ and $\R$-$\D$ links as
\begin{align}
\gamma_{\SR}[n]&\triangleq \frac{\gamma_{0}}{H^2+x^2[n]},
\gamma_{\RD}[n]\triangleq \frac{\gamma_{0}}{H^2+(D-x[n])^2}, \forall n. \label{eq:gammaRD}
\end{align}
The throughput maximization problem can  be formulated as
\begin{align}
 & \mathrm{(P1):} \  \underset{\substack{\{p_s[n]\}_{n=1}^{N-1}, \\ \{p_r[n]\}_{n=2}^N}}{\max}   \ \sum_{n=2}^N \log_2\left(1+p_r[n]\gamma_{\RD}[n] \right) \notag \\
  \text{s.t.}  &\  \sum_{i=2}^n \log_2\left(1+p_r[i]\gamma_{\RD}[i]\right) \leq \sum_{i=1}^{n-1} \log_2\left(1+p_s[i]\gamma_{\SR}[i]\right),\notag \\
   &\hspace{30ex} n=2,\cdots, N,\label{eq:InfoCausal2} \\
& \sum_{n=1}^{N-1} p_s[n] \leq E_s,\  \sum_{n=2}^N p_r[n] \leq E_r, \label{eq:PRConstr}  \\
& p_s[n] \geq 0, \ n=1,...,N-1, \\
&  p_r[n] \geq 0, \ n=2,...,N,
\end{align}
where \eqref{eq:PRConstr} corresponds to the average power constraints, with $E_s/N$ and $E_r/N$ being the average transmission power limits at $\Sn$ and $\R$, respectively. 
Denote the optimal value of (P1) as $R^\star$. The end-to-end throughput in bps/Hz is then given by $\tau^\star=R^\star/N$. 

\section{Optimal Solution}\label{sec:optSol}
 (P1) is a non-convex optimization problem due to the non-convex constraint \eqref{eq:InfoCausal2}. However, by introducing the slack variables $\{R_r[n]\}_{n=2}^N$, it can be equivalently written as
 \begin{align}
 \hspace{-5ex} & \mathrm{(P2):} \  \underset{\substack{\{p_s[n]\}_{n=1}^{N-1}, \\ \{p_r[n], R_r[n]\}_{n=2}^N}}{\max}   \ \sum_{n=2}^N R_r[n] \notag \\
  \text{s.t.}  &  \sum_{i=2}^n R_r[i] \leq \sum_{i=1}^{n-1} \log_2\left(1+p_s[i]\gamma_{\SR}[i]\right), n=2,\cdots,N\label{eq:InfoCausalConstr} \\
  &\ R_r[n] \leq \log_2\left(1+p_r[n]\gamma_{\RD}[n]\right),  n=2,\cdots,N \label{eq:RrConstr}\\
&\ \sum_{n=1}^{N-1} p_s[n] \leq E_s,\  \sum_{n=2}^N p_r[n] \leq E_r, \label{eq:PRConstr2} \\
&\ p_s[n] \geq 0, \ n=1,...,N-1, \label{eq:psConstr} \\
& \ p_r[n] \geq 0, \ n=2,...,N. \label{eq:prConstr}
\end{align}
If, at the optimal solution to (P2), there exists an $n'$ such that the constraint in \eqref{eq:RrConstr} is satisfied with strict inequality, we can always reduce the corresponding power $p_r[n']$ to make \eqref{eq:RrConstr} active, yet without decreasing the objective value of (P2). Thus, there always exists an optimal solution to (P2) such that all constraints in \eqref{eq:RrConstr} are satisfied with equality. As a result, (P2) is equivalent to (P1).  Note that (P2) is a convex optimization problem, which can be numerically solved by standard convex optimization techniques, such as the interior-point method \cite{202}. However, by applying the Lagrangian dual method, the structural properties of the optimal solution to $\mathrm{(P2)}$ can be obtained, based on which new insight can be drawn.

It can be verified that (P2) satisfies Slater's condition, thus, strong duality holds and its optimal solution can be obtained via solving  the dual problem \cite{202}. Furthermore, the power and rate allocations for $\Sn$ and $\R$ in (P2) are only coupled  via the information-causality
 constraint \eqref{eq:InfoCausalConstr}, which can be decoupled by studying its partial Lagrangian associated with this constraint.  Let $\lambda_n\geq 0$, $n=2,\cdots, N$, be the Lagrange dual variables corresponding to \eqref{eq:InfoCausalConstr}. The partial Lagrangian of (P2) can then be expressed as
\begin{align}
L &\left(\{p_s[n]\}, \{p_r[n], R_r[n], \lambda_n\} \right) \notag \\
=& \sum_{n=2}^N R_r[n] + \sum_{n=2}^N \lambda_n \left( \sum_{i=1}^{n-1} \log_2\left(1+p_s[i]\gamma_{\SR}[i]\right) -\sum_{i=2}^n R_r[i]\right)\notag \\
=& \sum_{n=2}^N \nu_n R_r[n] + \sum_{n=1}^{N-1}\beta_n \log_2\left(1+p_s[n]\gamma_{\SR}[n] \right),\label{eq:PartialL}
\end{align}
\begin{align}
\text{where} \hspace{10ex} &\beta_n \triangleq \sum_{i=n+1}^N \lambda_i, \ n=1,\cdots, N-1, \label{eq:betan}\\
&\nu_n\triangleq 1-\sum_{i=n}^N \lambda_i, \ n=2,\cdots, N. \label{eq:nun}
\end{align}

The Lagrange dual function of (P2) is then defined as
\begin{align}
g\left( \{\lambda_n\}\right) =
\begin{cases}
  \underset{\substack{\{p_s[n]\}_{n=1}^{N-1}, \\ \{p_r[n], R_r[n]\}_{n=2}^N}}{\max}  & \hspace{-2ex}  L \left(\{p_s[n]\}, \{p_r[n], R_r[n], \lambda_n\} \right) \notag \\
 \hspace{5ex} \text{s. t.}   & \hspace{-5ex} \eqref{eq:RrConstr}, \eqref{eq:PRConstr2}, \eqref{eq:psConstr}, \eqref{eq:prConstr}. \notag
\end{cases}
\end{align}
 The dual problem of (P2), denoted as (P2-D), is defined as $\min_{\lambda_n\geq 0, \forall n} g(\{\lambda_n\})$. Since (P2) can be solved equivalently by solving (P2-D), in the following, we first maximize the Lagrangian to obtain the dual function with fixed $\{\lambda_n\}$, and then find the optimal dual solutions $\{\lambda_n^\star\}$ to minimize the dual function. The optimal power and rate allocations at $\Sn$ and $\R$ are then obtained based on the dual optimal solution $\{\lambda_n^\star\}$.

Consider first the problem of maximizing the Lagrangian over $\{p_s[n]\}$ and $\{p_r[n], R_r[n]\}$ with fixed $\{\lambda_n\}$.
It follows from \eqref{eq:PartialL} that $g(\{\lambda_n\})$ can be decomposed as $g\left( \{\lambda_n\}\right)=g_s\left( \{\lambda_n\}\right)+g_r\left( \{\lambda_n\}\right)$, where
\begin{equation}
\begin{aligned}\label{eq:gs}
\hspace{-2ex} g_s\left( \{\lambda_n\}\right) =
\begin{cases}
\underset{\{p_s[n]\}}{\max} &  \sum_{n=1}^{N-1}\beta_n \log_2\left(1+p_s[n]\gamma_{\SR}[n] \right)   \\
 \text{s. t. }  & \sum_{n=1}^{N-1} p_s[n] \leq E_s, \\
  & \ p_s[n] \geq 0, \ n=1,...,N-1,
\end{cases}
\end{aligned}
\end{equation}
and
\begin{equation}
\begin{aligned}\label{eq:gr}
\hspace{-3ex} g_r\left( \{\lambda_n\}\right) =
\begin{cases}
\underset{\{p_r[n], R_r[n]\}}{\max} & \ \sum_{n=2}^{N}\nu_n R_r[n]  \\
 \text{s. t. }  &\hspace{-6ex} R_r[n] \leq \log_2\left(1+p_r[n]\gamma_{\RD}[n]\right), \forall n\\
 &\hspace{-6ex} \sum_{n=2}^{N} p_r[n] \leq E_r, \\
  &\hspace{-6ex} p_r[n] \geq 0, \ n=2,...,N.
  \end{cases}
\end{aligned}
\end{equation}
In other words, for any given dual variables $\{\lambda_n\}$, the optimal primal variables for Lagrangian maximization can be obtained by solving two parallel sub-problems \eqref{eq:gs} and \eqref{eq:gr} for $\Sn$ and $\R$, respectively. Note that both \eqref{eq:gs} and \eqref{eq:gr} are weighted sum-rate maximization problems each over $N-1$ parallel sub-channels, with the weights $\{\beta_n\}_{n=1}^{N-1}$ and $\{\nu_n\}_{n=2}^N$ determined by $\{\lambda_n\}_{n=2}^N$ given in \eqref{eq:betan} and \eqref{eq:nun}, respectively. Since $\lambda_n\geq 0$, $\forall n$, we have $\beta_n\geq 0$, $\forall n$, and $\{\beta_n\}_{n=1}^{N-1}$ and $\{\nu_n\}_{n=2}^N$ are {\it non-increasing} and {\it non-decreasing} over $n$, respectively. Furthermore, for problem \eqref{eq:gr} to have bounded optimal value, we must have $\nu_n\geq 0$, $\forall n$. To see this, suppose that there exists an $n'$ such that $\nu_{n'}<0$. Then problem \eqref{eq:gr} is unbounded when we let $R_r[n']=-t$, with $t\rightarrow \infty$. Since (P2) should have a bounded optimal value, it follows that the optimal primal and dual solutions of (P2) are obtained only when $\nu_n\geq 0$, $\forall n$, or equivalently $\sum_{n=2}^N \lambda_n \leq 1$.

By applying the standard Lagrange method and the Karush-Kuhn-Tucker (KKT) conditions, it is not difficult to conclude that the optimal solutions to \eqref{eq:gs} and \eqref{eq:gr} are respectively given by
\begin{align}
&p_s^\star[n]= \left[\eta \beta_n- \frac{1}{\gamma_{\SR}[n]} \right]^+, \ \forall n,\label{eq:ps} \\
& p_r^\star[n]= \left[ \xi \nu_n -\frac{1}{\gamma_{\RD}[n]}\right]^+, R_r^\star[n]= \left[\log_2 \left(\xi\nu_n \gamma_{\RD}[n] \right) \right]^+,  \forall n, \label{eq:pr}
\end{align}
where $\eta$ and $\xi$ are parameters ensuring $\sum_{n=1}^{N-1} p_s^\star[n]=E_s$ and $\sum_{n=2}^N p_r^\star[n]=E_r$, respectively, and $[a]^+\triangleq \max\{a,0\}$.

Next, we address how to solve the dual problem (P2-D) by minimizing the dual function $g(\{\lambda_n\})$ subject to $\lambda_n\geq 0$, $\forall n$, and the new constraint $\sum_{n=2}^N \lambda_n \leq 1$. This can be done by applying the subgradient-based method, e.g., the ellipsoid method \cite{200}. It can be shown that the subgradient  of $g(\{\lambda_n\})$ at point $\{\lambda_n\}$ is given by $\mathbf s=[s_2, \cdots , s_N]^T$, with $s_n=\sum_{i=1}^{n-1}\log_2\left(1+p_s^\star[i]\gamma_{\SR}[i]\right)-\sum_{i=2}^n R_r^\star[i]$, $\forall n$, where $\{p_s^\star[n]\}$ and $\{R_r^\star[n]\}$ are the solutions in \eqref{eq:ps} and \eqref{eq:pr} for the given $\{\lambda_n\}$. The procedures for finding the optimal dual solutions $\{\lambda_n^\star\}$ using the ellipsoid method are summarized in Algorithm~\ref{Algo:primDual} on the next page.

With the dual optimal solution $\{\lambda_n^\star\}$ to (P2-D) obtained, the primal optimal solution to (P2), denoted as $\{p_s^{\opt}[n]\}$ and $\{p_r^{\opt}[n], R_r^{\opt}[n]\}$, can be obtained by separately considering the following four cases.

{\it Case 1: $\beta_1^\star>0$ and $\nu_N^\star>0$,} which is equivalent to $\sum_{n=2}^N \lambda_n^\star>0$ and $\lambda_N^\star<1$. In this case, both the weighting vectors $\{\beta_n^\star\}$ for \eqref{eq:gs} and $\{\nu_n^\star\}$ in \eqref{eq:gr} have strictly positive components, and hence \eqref{eq:gs} and \eqref{eq:gr} are strict convex optimization problems and therefore have unique solution. As a result, the solution given in \eqref{eq:ps} and \eqref{eq:pr} corresponding to the  dual optimal variable  $\{\lambda_n^\star\}$ must be the primal optimal solution to (P2). Note that in this case, $\Sn$ and $\R$ both use up their maximum transmission power. Furthermore, \eqref{eq:ps} and \eqref{eq:pr} show that the optimal power allocations  across the different slots are given by the ``stair-case'' WF solution \cite{636}, with {\it non-increasing} and {\it non-decreasing} water levels at $\Sn$ and $\R$, respectively. Moreover, the water level changes after slot $n$ if and only if $\lambda_n^\star>0$, in which case, we have $\sum_{i=2}^n R_r^\opt[i]=\sum_{i=1}^{n-1} R_s^\opt[i]$ based on the complementary slackness condition, where $R_r^\opt[n]$ and $R_s^\opt[n]$ are the optimal transmission rate by $\R$ and $\Sn$ at slot $n$, respectively. In other words, all data stored in the buffer of $\R$ will be cleared after slot $n$ if $\lambda_n^\star>0$. 


{\it Case 2: $\beta_1^\star>0$ and $\nu_N^\star=0$,} or equivalently $\lambda^\star_N=1$ and $\lambda^\star_2=\cdots =\lambda^\star_{N-1}=0$. We then have $\beta_n^\star=1$, $\forall n$, and $\nu_n^\star=0$, $\forall n$. In this case, the weighted sum-rate maximization problem \eqref{eq:gs} reduces to sum-rate maximization problem, and its solution reduces to the classic WF power allocation with a constant water level \cite{209}, i.e., $p_s^\star[n]=\big[\eta-1/\gamma_{\SR}[n] \big]^+$, $\forall n$, with $\eta$ chosen such that $\sum_{n=1}^{N-1} p_s^\star[n]=E_s$. In this case, the unique Lagrangian maximizer $\{p_s^\star[n]\}$ must be the optimal power allocation for $\Sn$ corresponding to the primal optimal solution to (P2), i.e., $p_s^\opt[n]=p_s^\star[n]$, $\forall n$.
On the other hand, since $\nu_n^\star=0$, $\forall n$, problem \eqref{eq:gr} has non-unique solutions for Lagrangian maximization. The primal optimal solution can then be obtained by solving (P2) with the given optimal source power allocation $\{p_s^\opt[n]\}$. The resulting problem is a convex optimization problem of reduced complexity as compared to (P2).

Note that since $\lambda_N^\star=1$ for Case 2, the complementary slackness condition implies that $\sum_{n=2}^N R_r^\opt[n]=\sum_{n=1}^{N-1} R_s^\opt[n]$, i.e., the aggregated transmission rates at $\Sn$ and $\R$ are equal. Furthermore, as $\Sn$ (while not necessarily $\R$) must use up all its power to achieve such a rate balance, Case 2 corresponds to the scenario where the $\Sn$-$\R$ link is the bottleneck due to e.g., limited  power budget $E_s$ at $\Sn$ and/or poor channels $\{\gamma_{\SR}[n]\}$.

{\it Case 3: $\beta_1^\star=0$ and $\nu_N^\star>0$,} which corresponds to $\lambda_n^\star=0$, $\forall n$. Thus, we have $\beta_n^\star=0$, $\forall n$, and $\nu_n^\star=1$, $\forall n$. In this case, the optimal power allocation at $\R$ is given by the classic WF solution with a constant water level, i.e., $p_r^\opt[n]=\big[\xi-1/\gamma_{\RD}[n] \big]^+$, $\forall n$, with $\xi$ satisfying $\sum_{n=2}^{N} p_r^\star[n]=E_r$, and the resulting relay transmission rates are $R_r^\opt[n]= \left[\log_2 \left(\xi \gamma_{\RD}[n] \right) \right]^+$. On the other hand, as the source power allocation for Lagrangian maximization \eqref{eq:ps} is not unique, we may obtain the one as the primal optimal solution that minimizes the source transmission power while satisfying the information-causality constraint with the given relay transmission rates. The details are omitted for brevity.


{\it Case 4: $\beta_1^\star=0$ and $\nu_N^\star=0$}. This requires $\lambda_n^\star=0$, $\forall n$, on one hand, and also $\lambda_N^\star=1$ on the other hand. Thus, this case will not occur.

The complete algorithm for solving (P2) is summarized in Algorithm~\ref{Algo:primDual}.

\begin{algorithm}[H]
\caption{Algorithm for solving (P2)}\label{Algo:primDual}
\begin{algorithmic}[1]
\STATE Initialize $\lambda_n\geq 0$, $\forall n$, and $\sum_{n=2}^N \lambda_n\leq 1$.
\REPEAT
\STATE Obtain $\{p_s^\star[n]\}$ and $\{p_r^\star[n], R_r^\star[n]\}$ using \eqref{eq:ps} and \eqref{eq:pr}.
\STATE Compute the subgradient of $g(\{\lambda_n\})$.
\STATE Update $\{\lambda_n\}$ using the ellipsoid method subject to $\lambda_n\geq 0$, $\forall n$ and $\sum_{n=2}^N \lambda_n\leq 1$.
\UNTIL{$\{\lambda_n\}$ converges to the prescribed accuracy}.
\STATE Output $\{p_s^\opt[n]\}$ and $\{p_r^\opt[n], R_r^\opt[n]\}$ according to the three cases discussed.
\end{algorithmic}
\end{algorithm}

For the special case where the UAV moves unidirectionally towards $\D$, the optimal solution to (P2) can be obtained in closed-form. We first define the following functions. For any $0\leq \tilde E_s\leq E_s$, define a function $\bar R_s(\tilde E_s)\triangleq \sum_{n=1}^{N-1}\left[\log_2\left( \eta \gamma_{\SR}[n]\right) \right]^+$ as the aggregated rate transmitted by $\Sn$ using the classic WF power allocation with total transmission power $\tilde E_s$, and $\bar p_{s,n}(\tilde E_s)\triangleq \left[ \eta -1/\gamma_{\SR}[n]\right]^+$ as the corresponding power allocation for slot $n$, with $\eta$ satisfying $\sum_{n=1}^{N-1}\left[ \eta -1/\gamma_{\SR}[n]\right]^+=\tilde E_s$. Similarly, for $0\leq \tilde E_r \leq E_r$, define $\bar R_r(\tilde E_r)\triangleq \sum_{n=2}^N \left[\log_2\left(\xi \gamma_{\RD}[n]\right) \right]^+$, and $\bar p_{r,n}(\tilde E_r)\triangleq \left[\xi-1/\gamma_{\RD}[n] \right]^+$, with $\xi$ satisfying $\sum_{n=2}^N \left[ \xi-1/\gamma_{\RD}[n]\right]^+=\tilde E_r$. We then have the following result.

\begin{theorem}\label{theo:th1}
If $\gamma_{\SR}[n]$ is non-increasing over $n$ (correspondingly, $\gamma_{\RD}[n]$ is non-decreasing over $n$), an optimal power allocation to (P2) is
$p_s^\opt[n]=\bar p_{s,n} ( \tilde E_s^\opt), \ p_r^\opt[n]=\bar p_{r,n}( \tilde E_r^\opt), \ \forall n$,
\begin{align}
\text{where }
\big(\tilde E_s^\opt, \tilde E_r^\opt\big) =
\begin{cases}
\big(E_s, \hat E_r\big) \ & \text{ if }  \bar R_s(E_s) \leq \bar R_r(E_r) \\
 \big(\hat E_s,  E_r\big), \ & \text { otherwise},\notag
\end{cases}
\end{align}
with $\hat E_s$ and $\hat E_r$ denoting the unique solution to the equation $\bar R_s(\tilde E_s)=\bar R_r(E_r)$ and $\bar R_r(\tilde E_r)=\bar R_s(E_s)$, respectively. Furthermore, the corresponding optimal value of (P2) is $R^\opt = \min\{\bar R_s(E_s), \bar R_r(E_s)\}$.
\end{theorem}
\begin{IEEEproof}
Please refer to Appendix~\ref{A:th1}.
\end{IEEEproof}

Theorem~\ref{theo:th1} states that if the UAV moves unidirectionally towards $\D$, the optimal power allocations at both $\Sn$ and $\R$ reduce to the classic WF with constant water levels. Furthermore, the transmitter corresponding to the ``bottleneck'' link  would use up all its available power whereas the other transmitter reduces its power so as to balance the two links. Under such transmission strategies, the information-causality constraints are automatically guaranteed, which is intuitively understood since the $\Sn$-$\R$ link always has better channels, and hence higher power and rate, in earlier slots, whereas the reverse is true for the $\R$-$\D$ link.

\begin{figure*}
\centering
\includegraphics[scale=0.6]{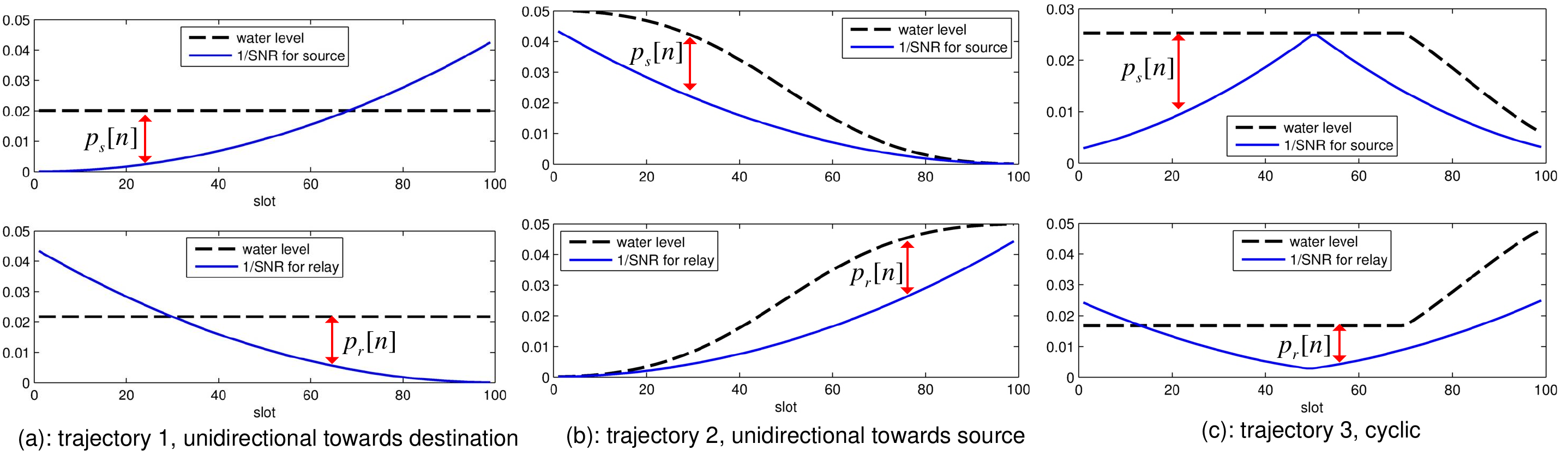}
\caption{Optimal power allocation for three different trajectories.\vspace{-3ex}}\label{F:PowerAllocations}
\end{figure*}

\section{Numerical Results}
In this section, numerical results are provided to compare the proposed mobile relaying versus the conventional static relaying techniques. We assume that   $\Sn$ and  $\D$ are separated by $D=2000$m. The system is operated at $5$GHz with $20$MHz bandwidth, and the noise power spectrum density is  $-169$dBm/Hz. Thus, the reference SNR at the distance $d_0=1$m can be obtained as $\gamma_{0}=80$dB. The average transmission power limits at both $\Sn$ and $\R$ are assumed to be $10$dBm. For both mobile and static relaying schemes, the altitude of the relays are fixed to be $H=100$m, and the maximum UAV speed is $\tilde{V}=50$m/s.

 Fig.~\ref{F:PowerAllocations} illustrates the optimal power allocations at $\Sn$ and $\R$ across different slots for mobile relaying with three specific UAV trajectories: (a) {\it unidirectional towards $\D$}, for which the UAV moves unidirectionally from $\Sn$ to $\D$ with the maximum speed; (b) {\it unidirectional towards $\Sn$}, where the UAV moves in the reverse direction from $\D$ to $\Sn$ with the maximum speed; (c) {\it cyclic between $D/4$ and $3D/4$}. It is observed from Fig.~\ref{F:PowerAllocations}(a) that for unidirectional movement to $\D$, the power allocations at both $\Sn$ and $\R$ follow the classic WF with a constant water level, which is in accordance with Theorem~\ref{theo:th1}; whereas for Fig.~\ref{F:PowerAllocations}(b) with the reverse movement, the water levels at $\Sn$ and $\R$ keep decreasing and increasing, respectively, which implies that the information-causality constraint is always active, i.e., the received data at $\R$ is immediately forwarded at the subsequent slot. For the cyclic movement shown in Fig.~\ref{F:PowerAllocations}(c), the water levels at both $\Sn$ and $\R$ are initially constant, and then decreases and increases respectively after certain period.

\begin{figure}
\centering
\includegraphics[scale=0.6]{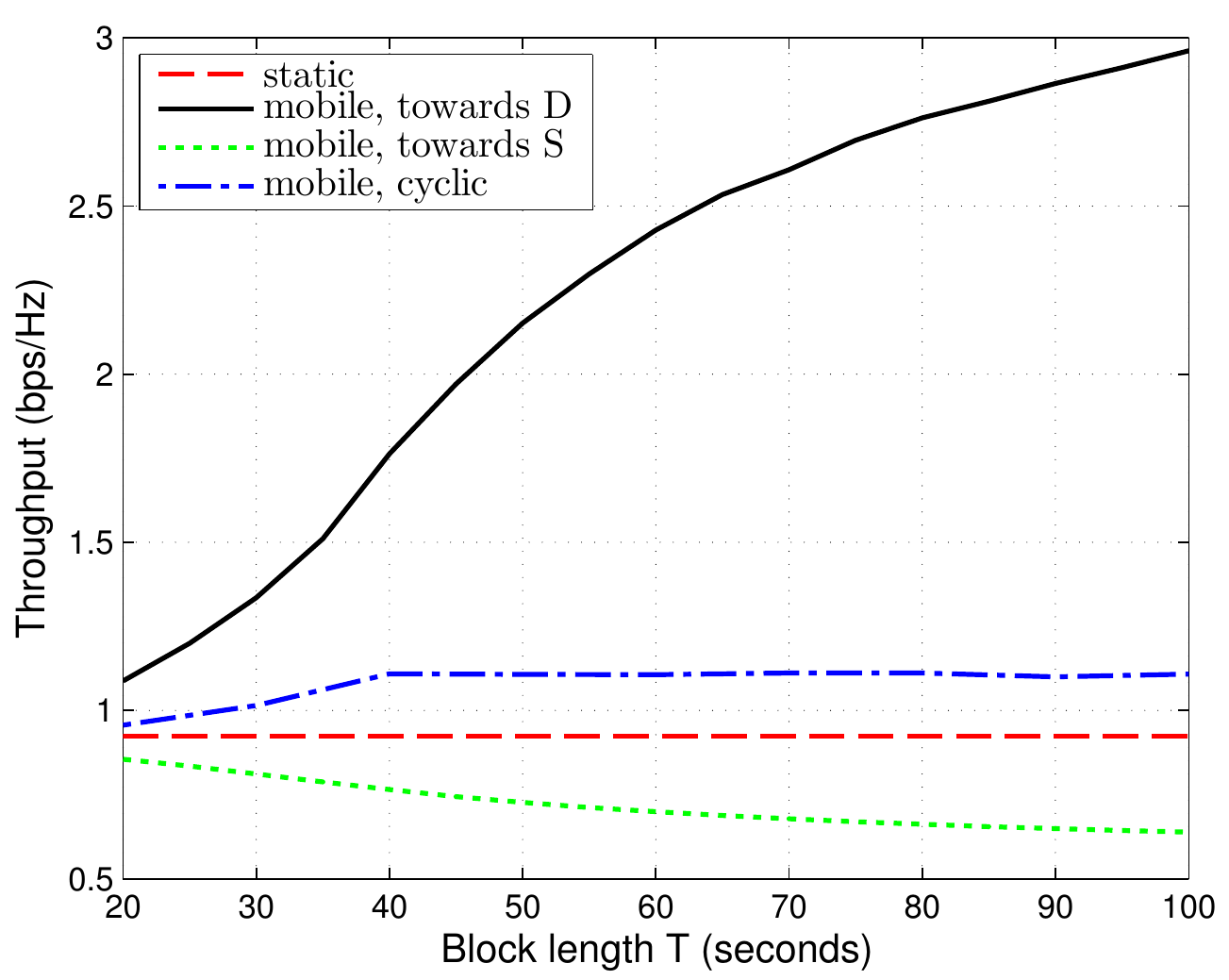}
\caption{Throughput comparison for different relaying schemes.\vspace{-3ex}}\label{F:ThroughputVSTFixedTrajectory}
\end{figure}

In Fig.~\ref{F:ThroughputVSTFixedTrajectory}, the throughput in bps/Hz versus the duration $T$ is plotted for the static versus mobile relaying with the three aforementioned mobility patterns. Note that when $T$ is sufficiently large, the UAV for the two unidirectional schemes could stay stationary above $\Sn$ (and above $\D$) for certain period before it moves towards $\D$ (after it arrives above $\D$). It is observed from the figure  that with the UAV moving unidirectionally towards $\D$, the mobile relaying scheme significantly outperforms the conventional static relaying, thanks to the reduced link distances for both information reception and forwarding by relay mobility from $\Sn$ to $\D$. In contrast, for unidirectional relay movement from $\D$ to $\Sn$, the performance is even worse than the conventional static relaying. This is expected since with this specific relay mobility pattern, both $\Sn$ and $\R$ are forced to allocate high power on weak channels due to the information-causality constraint, as can be seen from Fig.~\ref{F:PowerAllocations}(b). Such results imply the necessity of joint UAV trajectory and power allocations in order to realize the full benefit of mobile relaying technique.

\section{Conclusions}
This paper studies a new mobile relaying technique with high-mobility relays. By exploiting the predictable channel variations caused by relay mobility, the end-to-end throughput is maximized via dynamic power and rate allocations subject to a new {\it information-causality} constraint. It is shown that the optimal power allocations in general follow a ``stair-case'' WF structure with non-increasing and non-decreasing water levels at the source and relay, respectively. For the special case where the relay moves unidirectionally towards $\D$, the optimal power allocations reduce to the classic WF with constant water levels. Numerical results show that compared with the conventional static relaying, a dramatic throughput gain is achievable by the proposed mobile relaying, provided that the relay trajectory is appropriately designed. The joint optimization of trajectory design and resource allocations for mobile relaying systems will be pursued in our future work.

\bibliographystyle{IEEEtran}
\bibliography{IEEEabrv,IEEEfull}

\appendices
\section{Proof of Theorem~\ref{theo:th1}}\label{A:th1}
To show Theorem~\ref{theo:th1}, we need the following result.
\begin{lemma}\label{lemma:th1}
If $\gamma_{\SR}[n]$ is non-increasing over $n$, the  dual optimal solution $\{\lambda^\star_n\}$ must satisfy $\lambda_n^\star=0$, $\forall n=2,\cdots, N-1$.
\end{lemma}
\begin{IEEEproof}
We show Lemma~\ref{theo:th1} by contradiction. Suppose, on the contrary that there exists $2 \leq n' \leq N-1$ such that $\lambda^\star_{n'}>0$. Then this must correspond to Case 1 as discussed in Section~\ref{sec:optSol}. 
Thus, the transmission rates at $\Sn$ and $\R$ corresponding to the primal optimal solution of (P2) can be expressed as
\begin{align}
&R^\opt_s[n]= \left[\log_2\left(\eta\beta_n^\star \gamma_{\SR}[n] \right) \right]^+, \ n=1,\cdots, N-1, \label{eq:RsStar} \\
&R^\opt_r[n]= \left[\log_2\left(\xi \nu_n^\star \gamma_{\RD}[n] \right) \right]^+, \ n=2,\cdots, N. \label{eq:RrStar}
\end{align}
 Since both $\{\beta_n^\star\}$ and $\{\gamma_{\SR}[n]\}$ are non-increasing over $n$, it follows from \eqref{eq:RsStar} that $R^\opt_s[n]$ is non-increasing over $n$ too.  We thus have
$R_s^\opt[1] \geq R_s^\opt[2] \geq \cdots  \geq R_s^\opt[n'-1]$, which implies
\begin{align}
\sum_{n=1}^{n'-1} R_s^\opt[n] \geq (n'-1) R_s^\opt[n'-1]. \label{eq:rl2}
\end{align}
On the other hand, the non-increasing of $\gamma_{\SR}[n]$ implies that $\gamma_{\RD}[n]$ is non-decreasing, as can be inferred from \eqref{eq:gammaRD}. Together with the fact that $\nu_n^\star$ is non-decreasing, it follows from \eqref{eq:RrStar} that $R_r^\opt[n]$ is non-decreasing over $n$, or $R_r^\opt[2] \leq R_r^\opt[3] \leq \cdots  \leq R_r^\opt[n']$, which leads to
\begin{align}
\sum_{n=2}^{n'} R_r^\opt[n] \leq (n'-1) R_r^\opt[n'].\label{eq:rl3}
\end{align}
Furthermore, by applying the complementary slackness condition for primal and dual optimal solutions, the assumption  $\lambda^\star_{n'}>0$ implies that the information-causality constraint at slot $n'$ must be active, i.e.,
\begin{align}
\sum_{n=1}^{n'-1} R_s^\opt[n] =\sum_{n=2}^{n'} R_r^\opt[n]. \label{eq:rl1}
\end{align}
The relations \eqref{eq:rl2}-\eqref{eq:rl1} lead to
\begin{align}
R_s^\opt[n'-1]\leq R_r^\opt[n'].\label{eq:rl7}
\end{align}

Now consider the slots from $n'$ to $N$. Based on the non-increasing property of $R_s^\opt[n]$, we have
\begin{align}
R_s^\opt[N-1] \leq \cdots \leq R_s^\opt[n'] < R_s^\opt[n'-1],\label{eq:rl4}
\end{align}
where 
the strict inequality is true since $\lambda_{n'}^\star>0$ implies $\beta^\star_{n'}<\beta^\star_{n'-1}$, as can be seen from \eqref{eq:betan}. Similarly, we have
\begin{align}
R_r^\opt[n'] < R_r^\opt[n'+1]\leq  \cdots \leq R_r^\opt[N]. \label{eq:rl5}
\end{align}
The relations  \eqref{eq:rl7}-\eqref{eq:rl5} jointly lead to
\begin{align}
\sum_{n=n'}^{N-1} R_s^\opt[n] < \sum_{n=n'+1}^N R_r^\opt[n].\label{eq:rl6}
\end{align}
By adding \eqref{eq:rl1} and \eqref{eq:rl6}, we have
$\sum_{n=1}^{N-1} R_s^\opt[n] < \sum_{n=2}^N R_s^\opt[n]$,
which obviously violates the information-causality constraint \eqref{eq:InfoCausalConstr} at slot $N$, and thus $\{R_s^\opt[n]\}$ and $\{R_r^\opt[n]\}$ given in \eqref{eq:RsStar} and \eqref{eq:RrStar} cannot be primal optimal to (P2), or equivalently $\{\lambda_n^\star\}$ with $\lambda_n'>0$ cannot be dual optimal. This completes the proof of Lemma~\ref{lemma:th1}.
\end{IEEEproof}

With Lemma~\ref{lemma:th1}, the optimal solution to (P2) must either correspond to Case 2 or Case 3 as discussed in Section~\ref{sec:optSol}.
First, we address how to obtain the primal optimal solution to (P2) by assuming that the  dual optimal solution corresponds to Case 2.
Based on the discussions presented in Section~\ref{sec:optSol}, the optimal power allocation at $\Sn$ in this case is given by the classic WF solution with full transmission power, and the corresponding source transmission rate can be expressed as $R_s^\opt[n]=\left[ \eta -1/\gamma_{\SR}[n]\right]^+$, $\forall n$, with $\eta$ denoting the water level. Furthermore, the optimal power and rate allocations at $\R$ can be obtained by solving (P2) with the the pre-determined $R_s^\opt[n]$, i.e.,
\begin{equation}\label{eq:case2}
\begin{aligned}
   &\underset{\{p_r[n], R_r[n]\}_{n=2}^N}{\max}   \ \sum_{n=2}^N R_r[n]  \\
  \text{s.t.} \ &  \sum_{i=2}^n R_r[i] \leq \sum_{i=1}^{n-1} R_s^\opt[i], \ \forall n, \\
  &\ R_r[n] \leq \log_2\left(1+p_r[n]\gamma_{\RD}[n]\right), \ \forall n, \\
& \sum_{n=2}^N p_r[n] \leq E_r, \ p_r[n] \geq 0, \ \forall n.
\end{aligned}
\end{equation}

To solve problem \eqref{eq:case2}, we first consider its relaxed problem by discarding the information-causality constraint from slot $2$ to slot $N-1$, i.e., by solving
\begin{equation}\label{eq:case2Relaxed}
\begin{aligned}
   &\underset{\{p_r[n], R_r[n]\}_{n=2}^N}{\max}   \ \sum_{n=2}^N R_r[n]  \\
  \text{s.t.} \ &  \sum_{n=2}^N R_r[n] \leq \sum_{n=1}^{N-1} R_s^\opt[n], \\
  &\ R_r[n] \leq \log_2\left(1+p_r[n]\gamma_{\RD}[n]\right), \ \forall n, \\
& \sum_{n=2}^N p_r[n] \leq E_R, \ p_r[n] \geq 0, \ \forall n.
\end{aligned}
\end{equation}

\begin{lemma}\label{lemma:lm1}
The optimal power allocation to problem \eqref{eq:case2Relaxed} is $p_r^\opt[n]=\bar p_{r,n}( \hat E_r)$, with $\bar p_{r,n}(\cdot)$ and $\hat E_r$ defined in Theorem~\ref{theo:th1}.
\end{lemma}
\begin{IEEEproof}
With the function $\bar R_{r}(\tilde E_r)$ for any $0\leq \tilde E_r\leq E_r$ defined in Theorem~\ref{theo:th1}, it is not difficult to see that problem \eqref{eq:case2Relaxed} is  equivalent to finding the optimal total transmission power $\tilde{E}_r$ at $\R$ via solving
\begin{equation}
\begin{aligned}
\underset{0\leq \tilde{E}_R \leq E_r}{\max} \ & \bar{R}_r(\tilde{E}_R), \quad
\text{s.t. }  \bar{R}_r(\tilde{E}_R)\leq \sum_{n=1}^{N-1} R_s^\opt[n].
\end{aligned}
\end{equation}
Using the fact that $\bar{R}_r(\tilde{E}_r)$ monotonically increases with $\tilde E_r$, the results in Lemma~\ref{lemma:lm1}  can be readily obtained.
\end{IEEEproof}

\begin{lemma}\label{lemma:lm3}
If $\gamma_{\SR}[n]$ is non-increasing over $n$, problems \eqref{eq:case2} and \eqref{eq:case2Relaxed} are equivalent.
\end{lemma}

\begin{IEEEproof}
Note that problem \eqref{eq:case2Relaxed} is a relaxation of \eqref{eq:case2}. Thus, if the optimal solution to \eqref{eq:case2Relaxed} given in Lemma~\ref{lemma:lm1} is feasible to problem \eqref{eq:case2}, then it must also be the optimal solution to \eqref{eq:case2}, and hence the two problems are equivalent. We show this by contradiction. 

Suppose, on the contrary, that the solution given in Lemma~\ref{lemma:lm1} is not feasible to problem \eqref{eq:case2}, i.e., the information-causality constraint is violated for some slot from $2$ to $N-1$. Then let $n'$ be the smallest value in $\{2,\cdots, N-1\}$ that violates the constraint, i.e., $n'$ is the slot such that $\sum_{i=2}^{n'} R_r^\opt[i]>\sum_{i=1}^{n'-1} R_s^\opt[i]$ and  $\sum_{i=2}^{n'-1} R_r^\opt[i]\leq \sum_{i=1}^{n'-2} R_s^\opt[i]$, where $R_r^\opt[i]$ denotes the transmission rate at $\R$ corresponding to the power allocation in Lemma~\ref{lemma:lm1}. Then we must have $R_r^\opt[n']>R_s^\opt[n'-1]$. Furthermore, since $\gamma_{\SR}[n]$ is non-increasing over $n$, we have $R_s^\opt[n]$ and $R_r^\opt[n]$  non-increasing and non-decreasing, respectively, which gives
\begin{align}
R_s^\star& [N-1]\leq \cdots \leq R_s^\star[n']\leq R_s^\star[n'-1] \notag \\
& <R_r^\star[n']\leq R_r^\star[n'+1]\cdots \leq R_r^\star[N]. \label{eq:Ineq}
\end{align}
The inequality in \eqref{eq:Ineq} implies that $\sum_{i=n'+1}^N R_r^\opt[i]>\sum_{i=n'}^{N-1}R_s^\opt[i]$. Together with the assumption $\sum_{i=2}^{n'} R_R[i]>\sum_{i=1}^{n'-1} R_S[i]$, we have $\sum_{i=2}^N R_r^\opt[i] > \sum_{i=1}^{N-1} R_s^\opt[i]$, which contradicts the fact that $\{R_r^\star[i]\}$ is optimal to problem \eqref{eq:case2Relaxed}. Thus, the solution  given in Lemma~\ref{lemma:lm1}  must be feasible, and hence also the optimal solution, to problem \eqref{eq:case2}. This completes the proof of Lemma~\ref{lemma:lm3}.
\end{IEEEproof}

Lemma~\ref{lemma:lm1} and Lemma~\ref{lemma:lm3} give the optimal power allocations corresponding to Case 2 as specified in Section~\ref{sec:optSol}, or for the case when $\bar R_s(E_s)\leq \bar R_r(E_r)$ as in Theorem~\ref{theo:th1}. For Case 3 with $\bar R_s(E_s)\geq \bar R_r^\opt(E_r)$, the optimal solution as presented in Theorem~\ref{theo:th1} can be similarly obtained. The details are omitted for brevity. This completes the proof of Theorem~\ref{theo:th1}.

\end{document}